\documentclass[twocolumn,showpacs,prr,superscriptaddress,a4paper,floatfix,nofootinbib]{revtex4-1}
\usepackage{amsmath}
\usepackage{amsfonts}
\usepackage{fancybox}
\usepackage{eepic}
\usepackage{xcolor}
\usepackage{times}
\usepackage{latexsym}
\usepackage{pifont}
\usepackage{graphicx}
\usepackage{epstopdf}
\usepackage{bm}
\usepackage{scrextend}
\usepackage{amsfonts}
\def\kk{\boldsymbol{k}}

\newcommand{\qsgw}{QS$GW$}
\newcommand{\qsgwl}{QS$G\hat{W}$}
\newcommand{\qub}{School of Mathematics and Physics, Queen's University Belfast, Belfast BT7 1NN, Northern Ireland, United Kingdom}

\begin{document}
\title{Many body theory beyond $GW$: towards a complete description of 2-body correlated propagation}
\author{Brian Cunningham}
\affiliation{\qub}

\begin{abstract}
Starting with Hedin's equations, simple expressions for the irreducible self-energy are derived. The derivation with vertex effects included in the self-energy results in a number of terms beyond $GW$ such as second-order screened exchange (the term that also gives rise to vertex/excitonic effects in the polarisation) and an infinite series describing correlations between the added-particle(removed-hole) and the excited electrons and holes: the screened $T$-matrix channels. Two-body correlated propagation is considered, with the third propagating freely, however, 3-body correlations are discussed and these can be added hierarchically to the method. For electron-hole propagation the reducible polarisation is calculated, which results in an expression for the self-energy that can be solved analytically without the need for widely used expensive numerical methods for frequency integration or having to adopt the plasmon-pole approximation. The method requires diagonalisation of the usual particle-hole matrix \textendash~derived from the Bethe-Salpter equation \textendash~and also a particle-particle$+$hole-hole matrix that has a similar structure; with the second-order exchange present in all channels. The method goes beyond current widely used methods that adopt the $GW$ approximation, whilst being $ab$ $initio$ in nature and requiring relatively modest extensions to existing functionality, e.g., it could be implemented in the sophisticated quasiparticle self-consistent framework. It is anticipated the method will significantly improve on the accuracy of current state-of-the-art calculations, improving our understanding of processes in simple and strongly-correlated systems, with the additional contributions demanding a similar cost to methods that include vertex effects in the polarisation only.  
\end{abstract}
\maketitle

%
%
%
\section{Introduction}
Understanding interactions between sub-atomic particles is crucial for progress in science, medicine and technology. The most widely used method for calculating materials properties is the Kohn-Sham formulation of Density Functional Theory (DFT)\cite{PhysRev.136.B864,kohnsham}. DFT is well established, computationally efficient and often produces accurate values for materials properties. However, DFT has many limitations. The most notable is the empirical-like choice of exchange-correlation functional that is often “cherry-picked” to suit the problem at hand. This leads to issues such as the neglection of the derivative discontinuity and, as a result, the notorious band-gap underestimation. This issue can be mitigated with generalized Kohn-Sham schemes.\cite{PhysRevB.53.3764} A major challenge, however, arises when applying DFT to strongly correlated systems like NiO,\cite{ferdi94,QSGWhat,NiODFT}, often necessitating extensions like DFT+U\cite{PhysRevB.44.943} or DFT+DMFT\cite{PhysRevB.82.045105}. Wavefunction approaches are an alternative that can produce highly accurate ground state quantities such as total energy, however, these are computationally demanding and only applicable to small systems, whilst requiring the calculation of more information than is needed or useful.\cite{kohn_lecture} 

Many-body (perturbation) theory (MBT/MBPT) approaches based on the work of Hedin\cite{hedin} are becoming more popular, owing to advances in theory and computation and are the current state-of-the-art for excited states and spectral properties. Most implementations rely on the $GW$ approximation\cite{louie,GW_aryasetiawan,hedinGW} to the self-energy. Although an improvement on DFT, $GW$ tends to produce accurate values for, e.g., band gaps, sometimes as a result of fortuitous error cancellation\cite{Acharya21a,QSGWhat}, which is often overlooked. As we progress in theoretical developments, it is important that we do not become too concerned with blindly correcting quantities such as band gaps, and focus more on the important physical processes we can describe within a particular theory. An advanced theory that can give physical insight into complex phenomena in exotic systems may be of more use than a less advanced method that produces a better value for e.g., the band-gap, through luck. $GW$ also has numerous inadequacies, such as its starting-point dependence (and even in the many iterative procedures \textendash~see, for e.g. Refs.~\cite{SC_ener,ferdi94} \textendash~employed to remove this dependence) as well as missing terms in the so-called diagrammatic series. As a result, $GW$ tends to break down for strongly correlated systems;\cite{sabatino23,QSGW_paper} $GW$ satellites are not accurate even in weakly correlated systems;\cite{PhysRevLett.107.166401,PhysRevLett.131.216401} and $GW$ is not very accurate for determining total energies.\cite{Miyake02,doi:10.1021/acs.jctc.3c01200} Self-consistent techniques\cite{GW_aryasetiawan,PhysRevB.82.045105,PhysRevB.98.155143} are employed to remove the Kohn-Sham starting point, and include methods that make use of partial self-consistentcy\cite{SC_BE,SC_ener,PhysRevB.74.045102}, or single-shot methods that remove the starting point dependence self consistently, such as the QS$GW$ method.\cite{QSGW_prl1,QSGW_PRL,QSGW_paper} QS$GW$ - and its extensions\cite{QSGWhat} - have become popular in recent years. These methods determine an optimum starting point Hamiltonian, see, for e.g.,  Ref.~\cite{QSGW_paper}. 

Most $GW$ methods ignored the vertex that appears in Hedin's equations, however, methods to include vertex effects have become more common in recent years. Vertex corrections to the dielectric screening have been considered through an effective nonlocal static kernel constructed within time-dependent functional theory\cite{PhysRevB.81.085213,PhysRevLett.99.246403,PhysRevLett.94.186402}, and more recently we included these in the electron-hole polarisation in a QS$GW$ formalism\cite{QSGWhat} through solving the Bethe-Salpeter equation (BSE)\cite{BSE_paper,PhysRevLett.91.056402,PhysRevLett.91.256402} for the full polarisation matrix. The QS$G\hat{W}$ method in Ref.~\cite{QSGWhat} neglected the vertex in the expression for the self-energy $\Sigma=iGW\Gamma$. The method relies on the Ward identity\cite{QSGW_paper} that allows the cancellation between the so-called $Z$-factor (that effectively scales the non-interacting Green's function) and the vertex in the $q\rightarrow 0$, $\omega\rightarrow 0$ limit (the dominant part of $W$\cite{QSGW_paper}) and hybrid self-energies were also considered to mimic the effects of the missing vertex in the self-energy (and the missing electron-phonon interaction). Methods exist for inclusion of vertex effects beyond $GW$\cite{vert1,doi:10.1063/1.3249965,vert3,PhysRevLett.112.096401,PhysRevB.103.L161104}, however, some still rely on the Ward identity in the long range limit. Kutepov\cite{PhysRevB.94.155101,PhysRevB.95.195120} recently developed a scheme, similar to the QS$GW$ formalism of Ref.~\cite{QSGW_paper} that includes vertex effects in both the polarisation and the self-energy. CuCl\cite{QSGWhat,kutepov_cucl} is one such example of a system where vertex effects in the self-energy are important. This is attributed to the highest valence band being relatively flat and dispersionless (vertex effects in the polarisation were also found to have a stronger influence on flat bands in strongly correlated systems such as NiO\cite{QSGWhat}). Vertex effects were included in the multichannel Dyson equation method of Ref.~\cite{PhysRevLett.131.216401} and also the method described in Ref.~\cite{PhysRevB.106.165129}, that examines the functional derivative of the self energy with respect to the Green's function, $\delta \Sigma/\delta G$. Recently, we also considered the $T$-matrix channels (that will be shown to be contained within the vertex) for a positron (the antiparticle of an electron)\cite{nature_positron,prl_pos} in finite systems, and these were found to be essential for the description of positron interactions with matter and were even essential to, e.g., enable positron binding in non-polar molecules. 

Simple analytic expressions for the energy-dependent self-energy are derived. The expressions contain interactions: bare and screened (that can be calculated at the level of the Random-Phase-Approximation (RPA), or in some kind of self-consistent manner, as in Ref.\cite{QSGWhat}); the energy eigenvalues of the single-particle Hamiltonian and also the eigenvectors and eigenvalues of two 2-body matrices (similar to the particle-hole matrix in Ref.~\cite{QSGWhat}). These expressions are derived from Hedin's equations with vertex effects in the polarization and self-energy (determined by calculating $\delta\Sigma/\delta G$, $ab$~$initio$). In this article we only keep terms that describe two-body correlations between the added particle (or removed hole) and the excited electrons and holes with the third particle propagating independenetly. For example, when the added particle interacts with an excited electron, the hole will propagate independently, and for this particular diagram RPA and excitonic effects will be neglected (they do still appear in the $GW$ diagram, where the added particle propagates independently). The supression of multiple excitations (e.g., biexcitons) is discussed. The usual $GW$ diagrams (RPA+excitonic effects) appear on an equal footing with the $T$-matrix channels (shown to be contained in Hedin's equations, as in Ref.~\cite{PhysRevB.106.165129}) that include particle-particle (and hole-hole) correlations, with second order screened exchange\cite{SEX} appearing throughout the infinte series. The screening $W$ in the expression $\Sigma=iGW\Gamma$ is replaced with the reducible polarization $\Pi$, where $W=v+v\Pi v$, and this is calculated instead of the irreducible polarisation, allowing the frequency integration to be performed analytically without the need for expensive numerical frequency integration (as in Ref.~\cite{QSGW_paper}). Any appreance of $W$ in the derivation then comes from $\delta\Sigma/\delta G$ and these are assumed static when arriving at the final expression for the self energy. The expression for the vertex is recursive and requires truncation (as discussed in Ref.~\cite{PhysRevB.106.165129}), however, with the inclusion of higher order terms in this series for the added-particle excited-hole diagram, it is shown that the eigenvectors and eigenvalues of two 2-body matrices are required, rather than three. The method produces the usual $GW$ diagram (that can also include vertex effects in $W$, as in Ref.~\cite{QSGWhat}) and as we will see, the second-order screened exchange diagram can be included with little extra computation required. The $T$-matrix diagrams have similar structure and should require similar resources. However, the size of, e.g., the particle-particle matrix, will be different to the particle-hole matrix and may often increase the computational cost, as in Ref.~\cite{nature_positron}. The supression of self-polarization effects (more pronounced in strongly correlated systems) is also apparent. The method is general, intuitive and could be implemented in most electronic structure software packages, with the possibility of adding three-body correlations\cite{stef} (illustrated throughout this paper) hierarchically.

\par
\section{Derivation of the Self-energy}
In this method, the following set of closed coupled equations\cite{PhysRev.139.A796,schwinger,schwinger2} of Hedin are to be solved iteratively:
%
\begin{align}
  \Sigma(1,2)&={\rm i}\int{\rm d}(34)~G(1,3^{+})W(4,1)\Gamma(3,2;4)\label{eq:selfE1}\\
  G(1,2)&=G_{0}(1,2) + \int{\rm d}(34)~G_{0}(1,3)\Sigma(3,4)G(4,2)\label{eq:GreenF}\\
 W(1,2)&=v(1,2)+\int{\rm d}(34)v(1,3)P(3,4)W(4,2)\label{eq:scrpot}\\
P(12)&=-{\rm i}\int{\rm d}(34)G(1,3)G(4,1^{+})\Gamma(3,4;2)\label{eq:irrpol}\\
 \Gamma(1,2;3)&=\delta(1,2)\delta(1,3) + \nonumber \\&~~~\int {\rm d}(4567)\frac{\delta \Sigma(1,2)}{\delta G(4,5)} G(4,6) G(7,5)\Gamma(6,7;3)\label{eq:vert}\end{align}
where the indices are composite indices that subsume space, time and possibly spin: $(1)=(\bm{r}_1,t_1,\sigma_1)$, $G$ is the Green's function ($G_0$ is the non-interacting $G$), $v(\boldsymbol{r},\boldsymbol{r}')=1/|\boldsymbol{r}-\boldsymbol{r}'|$ is the bare Coulomb interaction, $W$ is the screened Coulomb interaction, $\Gamma$ is the irreducible vertex function, $P$ is the irreducible polarizability (the functional derivative of the induced density with respect to the total potential) and $\Sigma$ is the self-energy operator. In Eqs.~\eqref{eq:selfE1} and~\eqref{eq:irrpol}, the $+$ superscript implies $t'=t+\eta$, with $\eta\rightarrow 0^{+}$ and this will now be dropped for simplicity. From here on, repeated indices appearing on one side of an equality implies integration/summation. Making use of these equations, we can write the self-energy as
\begin{equation}
\Sigma(1,2)=\Sigma^{GW}(1,2)+iG(1,3)W(4,1)i\frac{\delta\Sigma(3,2)}{\delta G(5,6)}P(5,6;4),
\end{equation}
where $P(5,6;4)=-iG(5,7)G(8,6)\Gamma(7,8;4)$ is the 3-point extension\footnote{Quantities can be 2, 3 or even 4-point functions. 2 or 3 point functions are contractions of the full 4-point function, i.e., $P(1,2)=P(1,1;2,2)$ and $P(1,2;3)=P(1,2;3,3)$.} of Eq.~(\ref{eq:irrpol}). Introducing the reducible polarisation:
\begin{align}
\Pi(1,2)=P(1,2)+P(1,3)v(3,4)\Pi(4,2)\end{align} such that\begin{align}
W(1,2)=v(1,2)+v(1,3)\Pi(3,4)v(4,2)\label{eq:redPi}
\end{align}
we can write the self energy as
\begin{equation}\begin{array}{rl}
\Sigma(1,2)=&\Sigma^{ex}(1,2)+iG(1,2)v(2,3)\Pi(3,4)v(4,1)+\\
&iG(1,3)v(4,1)i\displaystyle{\frac{\delta\Sigma(3,2)}{\delta G(5,6)}}P(5,6;4)\\
&iG(1,3)v(4,7)\Pi(7,8)v(8,1)i\displaystyle{\frac{\delta\Sigma(3,2)}{\delta G(5,6)}}P(5,6;4)
\end{array}
\end{equation}
with $\Sigma^{ex}$ the exchange contribution. Making use of the expression for the reducible polarisation we can write
\begin{equation}
\Sigma(1,2)=\Sigma^{ex}(1,2)+iG(1,3)\chi(3,2;5,6)\Pi(5,6;4)v(4,1),
\end{equation}
where $\chi(3,2;5,6)=v(2,5)\delta(3,2)\delta(5,6)+i\displaystyle{\frac{\delta\Sigma(3,2)}{\delta G(5,6)}}$ is the kernel that appears in the BSE (with $i\delta\Sigma/\delta G$ usually approximated as $-W$) for the reducible polarisation.\cite{kernPi} The $W$ that appears in the expression for the self-energy has been replaced with $v+v\Pi v$, and as a result, any appearance of $W$ now arises from the functional derivative of the self-energy with respect to the Green's function.

\begin{figure}[h!]
\includegraphics[width=0.3\textwidth,clip=true,trim=1.0cm 18.5cm 4.0cm 3.8cm]{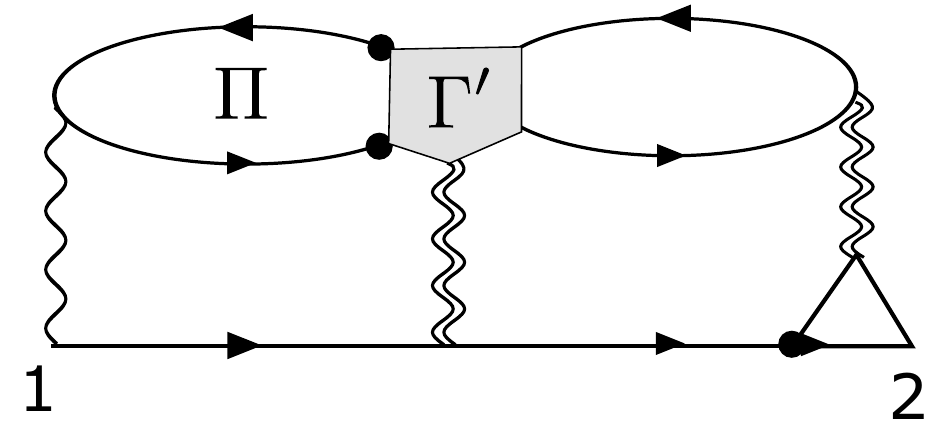}
\caption{Diagram corresponding to the fourth term on the R.H.S. of Eq.~(\ref{eq:dSigdG}). Single lines correspond to propagators (Green's functions), with bubbles corresponding to pairs of particle-hole propagators, with $\Pi$ the reducible polarisation describing the series of interacting electron-hole bubbles. Single(double) wavy lines are the bare(screened) Coulomb interaction and the vertex is such that the first index is indicated with a black circle and the second is indicated by the direction of the arrow at the circle. Higher order correlations are contained in the functional derivative of the vertex, $\Gamma'$.}
\label{fig:gammap}
\end{figure}

\begin{figure}[h!]
\includegraphics[width=0.26\textwidth,clip=true,trim=2.0cm 11.5cm 2.0cm 7cm]{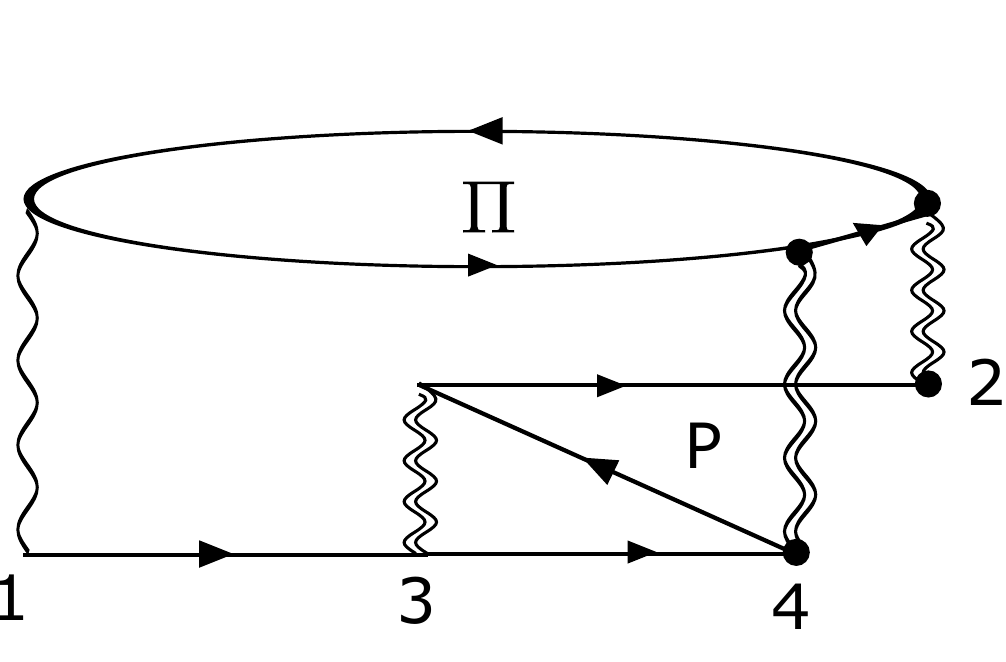}
\caption{A double excitation as a result of including $\delta/\delta G \left\{\delta\Sigma/\delta G\right\}$ (see text). Note the three point polarisations $P$ and $\Pi$ are such that the two black circles and arrows indicate the order of the indices: first is at a circle, with the propagator pointing towards the third and the other circle indicates the second.}
\label{fig:d2SigdG2}
\end{figure}

We will now focus on the functional derivative $i\delta\Sigma/\delta G$ (similar to Ref.~\cite{PhysRevB.106.165129}), resulting in
\begin{equation}\begin{array}{rl}
i\displaystyle\frac{\delta\Sigma(3,2)}{\delta G(5,6)}=&-W(8,3)\Gamma(6,2;8)\delta(3,5)\\& -G(3,7)\displaystyle\frac{\delta W(8,3)}{\delta G(5,6)}\Gamma(7,2;8)\\&-G(3,7)W(8,3)\displaystyle\frac{\delta\Gamma(7,2;8)}{\delta G(5,6)}.\end{array}\label{eq:dSig0}
\end{equation}
First, lets look at the functional derivative of $W$, making use of Eq.~(\ref{eq:scrpot}): $W'=vP'W+vPW'$, where for shorthand the $'$ implies the functional derivative. This can be rearranged to $W'=WP'W$, since $W=(1-vP)^{-1}v$ and thus
\begin{equation}\begin{array}{rl}
\displaystyle\frac{\delta W(8,3)}{\delta G(5,6)}= -i & W(8,9)\displaystyle\frac{\delta}{\delta G(5,6)}\left\{G(9,11)\times\right.\\&\left.G(12,9)\Gamma(11,12;10)\right\}W(10,3),
\end{array}\end{equation}
making use of Eq.~\ref{eq:irrpol}, and therefore\begin{widetext}
\begin{equation}\begin{array}{rl}
i\displaystyle\frac{\delta\Sigma(3,2)}{\delta G(5,6)}=&-W(8,3)\Gamma(6,2;8)\delta(3,5)+iG(3,7)W(8,5)G(12,5)\Gamma(6,12;10)W(10,3)\Gamma(7,2;8)\\
& +iG(3,7)W(8,6)G(6,11)\Gamma(11,5;10)W(10,3)\Gamma(7,2;8)\\&+iG(3,7)W(8,9)G(9,11)G(12,9)\displaystyle\frac{\delta\Gamma(11,12;10)}{\delta G(5,6)}W(10,3)\Gamma(7,2;8)-G(3,7)W(8,3)\displaystyle\frac{\delta\Gamma(7,2;8)}{\delta G(5,6)}.
\end{array}\label{eq:dSigdG}\end{equation}\end{widetext}
The two terms on the last line of Eq.~\ref{eq:dSigdG} contain the functional derivative of the vertex w.r.t. $G$. Figure~\ref{fig:gammap} presents the diagram corresponding to this first term. It is clear that this diagram/term will capture three-body correlated propagation (i.e., the excited electron and hole are now correlated, whilst one is already correlated with the added particle) and therefore neglected, since we are only considering two body correlated propagation.\footnote{The multichannel Dyson equation method of Ref.~\cite{PhysRevLett.131.216401} examines the 3-body Green's function for photoemission spectra} To calculate $\Gamma'$ the second functional derivative of the self energy is neglected (for now) and from Eq.~\ref{eq:vert}, $\delta\Gamma(7,2;8)/\delta G(5,6)=i\delta\Sigma(7,2)/\delta G(5',6')\delta P(5',6';8)/\delta G(5,6)$, giving 
\begin{widetext}
\begin{equation}\begin{array}{rl}
i\displaystyle\frac{\delta\Sigma(3,2)}{\delta G(5,6)}=&-W(8,3)\Gamma(6,2;8)\delta(3,5)+iG(3,7)W(8,5)G(12,5)\Gamma(6,12;10)W(10,3)\Gamma(7,2;8)\\
& +iG(3,7)W(8,6)G(6,11)\Gamma(11,5;10)W(10,3)\Gamma(7,2;8)
\\&-G(3,7)W(8,3)\displaystyle\frac{\delta\Sigma(7,2)}{\delta G(5,6')}G(8',6')\Gamma(6,8';8)
\\&-G(3,7)W(8,3)\displaystyle\frac{\delta\Sigma(7,2)}{\delta G(5',6)}G(5',7')\Gamma(7',5;8)+...,
\end{array}\end{equation}\end{widetext}
where the diagram in $\delta\Gamma/\delta G$ that contains $\delta\Gamma/\delta G$ (since the series are recursve) from $\delta P/\delta G$ is neglected as it can be seen that this corresponds to three-body correlated propagation, as in Fig.~\ref{fig:gammap}. Lets now take a look at the second functional derivative of the self-energy. If the first term in the expression for $\Sigma'$ (Eq.~\ref{eq:dSigdG}) is assumed to be the dominant term (as is usually done in the BSE for the polarisation) and the vertex is neglected (i.e., just look at the first term in an infinite series), it can be seen from Fig.~\ref{fig:d2SigdG2} (for the propagation of an added particle) that this diagram can describe an added particle that creates a correlated series of electron-hole pairs at time $t_1$, and another correlated electron-hole pair at time $t_3$ via a screened Coulomb interaction, then at time $t_4$ the added particle combines with the second hole, scattering the first excited electron, that propagates until combining with the first hole, allowing the second excited electron to be removed from the system. This diagram is neglected since it describes a double excitation (more than one electron-hole pair propagating at the same time). 

To include vertex effects in the polarisation, Eq.~\ref{eq:irrpol}, it is common to take just the first term in Eq.~\ref{eq:dSigdG} with $\Gamma=1$. Upon analysing the polarisation diagramatically it is clear that the other terms in Eq.~\ref{eq:dSigdG} also gives rise to, e.g., the creation of electron-hole pairs from an excited electron or hole, resulting in the simultaneous propagation of multiple electron-hole pairs. It is discussed in Ref.~\cite{PhysRevLett.131.216401} how that method could be extended to couple 2-particle Green's functions and 4-particle Green's functions to describe double excitations and biexcitons. These processes are neglected in this work. 
\begin{figure}[h!]
\includegraphics[width=0.48\textwidth,clip=true,trim=0.0cm 0.0cm 0.0cm 0.0cm]{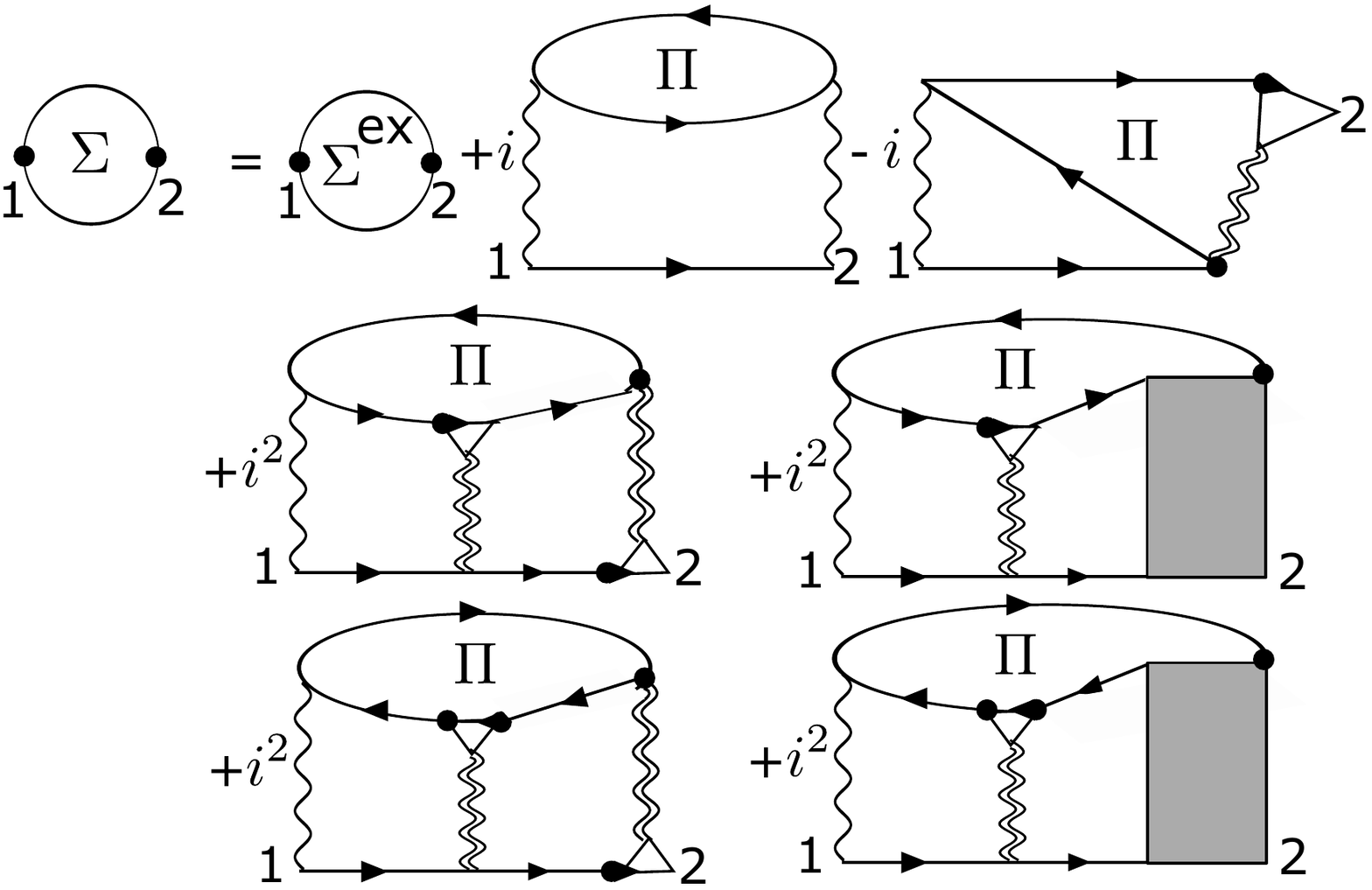}
\caption{Truncated series of diagrams considered in the method. The first term is the Fock exchange, then the $GW$ diagram, followed by the second order exchange diagram. Then we have particle-particle/hole-hole and particle-hole channels (with the added/removed particle), with the shaded rectangle equal to $i\delta\Sigma/\delta G$.}
\label{fig:7diagrams}
\end{figure}

\begin{figure}[h!]
\includegraphics[width=0.48\textwidth,clip=true,trim=0.0cm 9.0cm 1.0cm 15.0cm]{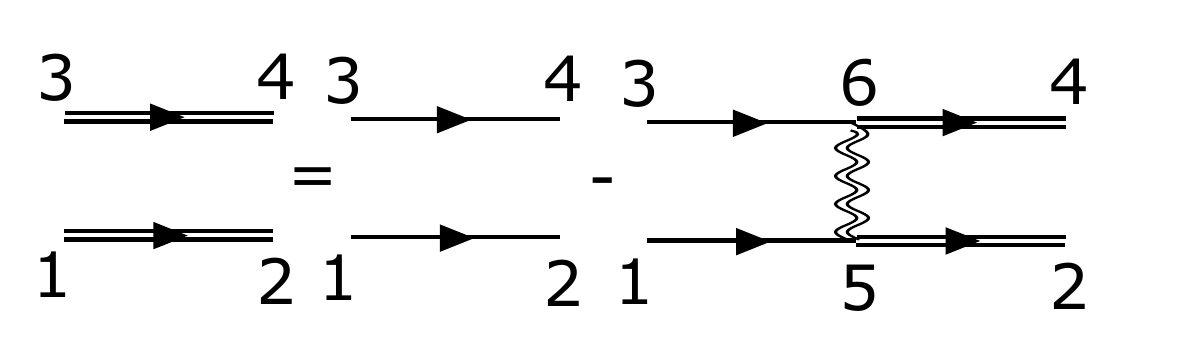}
\caption{The correlated particle-particle propagator $L(1,3;2,4)=L_0(1,3;2,4)-L_0(1,3;5,6)W(6,5)L(5,6;2,4)$, with $L_0(1,3;2,4)=-iG(1,2)G(3,4)$.}
\label{fig:L}
\end{figure}

Fig.~\ref{fig:7diagrams} presents the truncated series of diagrams that must now be evaluated. 
Considering the fifth and seventh diagrams on the R.H.S. and inserting the expression for $i\delta\Sigma/\delta G$ it can be seen that an infinite ladder series between the added particle and the excited electrons/holes is obtained (analogous to the usual electron-hole polarisation), plus an exchange type of interaction (arising from the first term in Eq.~\ref{eq:dSigdG}, that corresponds to the third diagram in Fig.~\ref{fig:7diagrams}) on the right of each term in these series. For diagrams that describe correlated propagation between the added particle (or removed hole) and the excited electrons or holes, subsequent interactions with the other type of particle, i.e., the holes or electrons, are supressed. For the $T$-matrix diagrams, the added particle is correlated with either the excited electrons or the holes, not both. However, the added-electron(removed-hole) can always combine with the hole(excited-electron) via second-order screened exchange arising from the vertex. For the terms that describe correlations with the added particle it is assumed\footnote{Note that it is also possible to replace the $\Pi v$ on the left with $PW$ and instead assume $P=-iGG$.} that $\Pi=-iGG$ (suppressing excited electron-hole correlations, since it was stated that one of these will have to act as a spectator). The self-energy is then evaluated as follows:
\begin{align}
\Sigma(1,2)=&\Sigma^{ex}(1,2)+iG(1,3)\widetilde{\chi}(3,2;5,6)\Pi(5,6;4)v(4,1)\nonumber
\\&+iL_{pp}(1,4;3,6')v(4,1)G(5,4)\chi_{pp}(3,2;5,6')\nonumber\\
&+iL_{ph}(1,4;3,5')v(4,1)G(4,6)\chi_{ph}(3,2;5',6),
\end{align}  
where the $L$ are correlated propagators given by
\begin{align}
L_{pp}(1,2;3,4)=&-iG(1,3)G(2,4)+\left\{-iG(1,5)G(2,6)\right\}\times\nonumber\\
&~~~~~~~~~~~~~~~\left\{-W(6,5)\right\}L_{pp}(5,6;3,4)\nonumber\\
L_{ph}(1,2;3,4)=&-iG(1,3)G(4,2)+\left\{-iG(1,5)G(6,2)\right\}\times\nonumber\\
&~~~~~~~~~~~~~~~\left\{-W(6,5)\right\}L_{ph}(5,6;3,4)\label{eq:L}
\end{align}
and
\begin{align}
\widetilde{\chi}(3,2;5,6)=&v(2,5)\delta(3,2)\delta(5,6)-W(2,3)\delta(3,5)\delta(2,6)\nonumber\\
\chi_{mn}(3,2;5,6)=&W(2,5)\delta(3,2)\delta(5,6)-W(2,3)\delta(3,5)\delta(2,6)\label{eq:chi},
\end{align}
with $mn$ equal to $pp$ or $ph$. The $\chi$ have a direct and an exchange-like contribution. The two-body correlated particle-particle propagator $L$ is illustrated in Fig.~\ref{fig:L}, with the added particle-hole propagator being obvious. Note that in Eqs~\ref{eq:L} and \ref{eq:chi} the vertex attached to the screened interactions is dropped for now (effectively truncating the infinite series for $\Gamma$). Inclusion of the vertices everywhere would make the problem effectively impossible to solve in practice. It could be argued that these vertices can be dropped here by making use of the Ward identity.\cite{QSGW_paper,PhysRevA.90.032506} The Ward identity, derived from Gauge invariance, has been used in the past to approximate the vertex in the calculation of the self-energy,\cite{QSGW_paper} and more recently to mimic the long-range, static ($\bf{q},\omega\rightarrow 0$) contribution from the vertex only.\cite{fxc_kernel,PhysRevB.103.L161104,prb_new} In this limit, $\Gamma\rightarrow 1-\partial \Sigma/\partial \omega=1/Z$, where $Z$ is the $Z$-factor renormalization $G\approx ZG_0$, as discussed in Ref.~\cite{QSGW_paper}. Vertex effects are included in this method here, but since the equations are recursive, vertices appear in the series that result and these series will need truncated (as discussed in Ref.~\cite{PhysRevB.106.165129}, where initially only the first non-trivial step in the resummation of the expression for the vertex is taken). It can be argued that when the series is truncated by setting $\Gamma=1$ an interacting Green's function connected to this vertex should be replaced with a non-interacting one, making the method more accesible without sacrificing too much on the numerical accuracy that is expected. This is discussed further in the Conclusions. Note that the propagators $L$ contain zeroth order diagrams that describe three non-interacting particles. These diagrams, however, are already contained in $GW$ and second-order exchange diagram, and so care will need to be taken to ensure each term is only included once (the zeroth order contributions will need to be subtracted).

\begin{figure}[h!]
\includegraphics[width=0.48\textwidth,clip=true,trim=0.0cm 9.0cm 0.5cm 15.0cm]{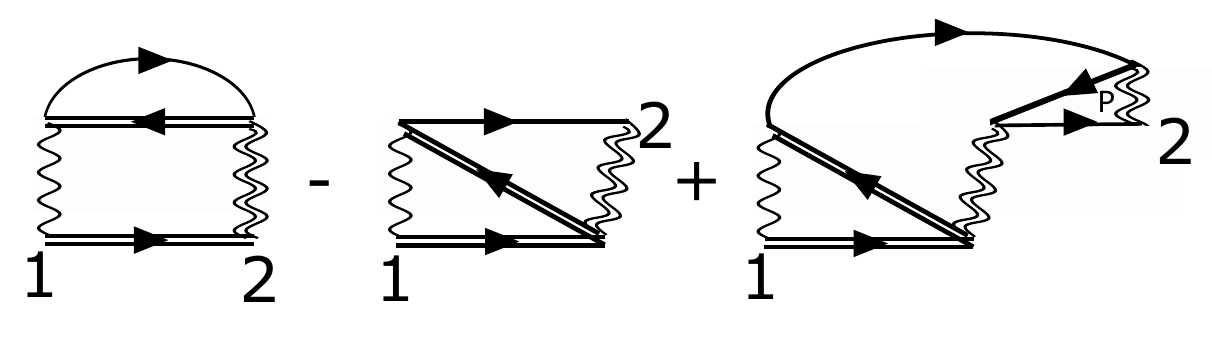}
\caption{Correlated propagation between the added electron and the hole. The third term arises from considering the contribution $i\Sigma' P$ from the vertex that was attached to the screening in the second diagram (see text).}
\label{fig:ppPi}
\end{figure}

The contributions to the self-energy contains the particle-hole reducible polarisation $\Pi$ for the correlated excited electron-hole diagram and the irreducible polarisation $P$ for the added particle-hole diagram. To evaluate would require the diagonalisation of a large matrix (see below) for both terms. However, this can be overcome by considering the added particle-hole channel and for the exchange diagrams including the vertex attached to the screening explicitly (as in the third diagram on the right of Fig.~\ref{fig:7diagrams}). The three terms that arise for this channel are then illustrated in Fig.~\ref{fig:ppPi}. The third term arises from taking the first term in Eq.~\ref{eq:dSig0} for $i\delta\Sigma/\delta G$ and just taking the first term in the recursive series for $\Gamma$ that results. The first and third terms in Fig.~\ref{fig:ppPi} can then be combined since $P+PWP=\Pi$ and from the second term we have $PW=\Pi v$. Therefore the added particle-hole propagator can be replaced with the 4-point extension of the usual reducible polarisation that contains an RPA-like series of bubbles with a ladder series of interactions within.

Lets now assume the widely used approximation that all interactions (except the $W$ in $iGW\Gamma$ that we replaced with $v+v\Pi v$) are instantaneous and that quantities are only dependent on time differences $t=t_1-t_2$. Transforming to a basis of single particle states with $\Sigma_{mn}(t)=\int\int\psi_m^{*}(\boldsymbol{r})\Sigma(\boldsymbol{r},\boldsymbol{r}';t)\psi_{n}(\boldsymbol{r}')d\boldsymbol{r}d\boldsymbol{r}'$ gives
\begin{align}
\Sigma_{mn}(t)=\Sigma_{mn}^{ex}(t)+&i\hspace{-0.2cm}\sum_{n'\hspace{-0.1cm}\scriptsize\begin{array}{l}n_1n_2\\n_3n_4\end{array}}\hspace{-0.2cm}\displaystyle\left\{\widetilde{\chi}_{\scriptsize\begin{array}{l}n'n\\n_1n_2\end{array}}\hspace{-0.2cm}\displaystyle{\Pi_{\scriptsize\begin{array}{l}n_1n_2\\n_3n_4\end{array}}(-t)G_{n'}(t)V_{\scriptsize\begin{array}{l}n_3n_4\\n'm\end{array}}}\right.\nonumber\\
&+\hspace{-0.0cm}\chi^{pp}_{\scriptsize\begin{array}{l}n_3n\\n'n_4\end{array}}\hspace{-0.1cm}\displaystyle{L^{pp}_{\scriptsize\begin{array}{l}n_1n_2\\n_3n_4\end{array}}(t)G_{n'}(-t)V_{\scriptsize\begin{array}{l}n'n_2\\n_1m\end{array}}}\nonumber\\
&+\left.\hspace{-0.0cm}\widetilde{\chi}^{ph}_{\scriptsize\begin{array}{l}n_2n\\n_1n'\end{array}}\hspace{-0.1cm}\displaystyle{\Pi_{\scriptsize\begin{array}{l}n_1n_2\\n_3n_4\end{array}}(-t)G_{n'}(t)V_{\scriptsize\begin{array}{l}n_3n'\\n_4m\end{array}}}\right\},
\label{eq:Sigma_4p}\end{align}
where $G_{n'}(t)=\pm \frac{1}{i} e^{-i\varepsilon_{n'}t}\theta(\pm t)$ with $+(-)$ for particles/holes and $\psi_{n'}$ and $\varepsilon_{n'}$ are the eigenfunctions and eigenvalues for the Hamiltonian corresponding to $G$.\footnote{If $H_0$ is the non-interacting Hamiltonian then $G=G_0$ is the non-interacting Green's function that is widely used in many body methods and this discussed further in the text.} $\widetilde{\chi}^{ph}$ is for the added particle-hole diagram (given in Eq.~\ref{eq:chi}) except now the exchange interaction is bare instead of screened as a result of the mechanism presented in Fig.~\ref{fig:ppPi}. Note that the indices are composite, containing band, spin and wavevector (for extended systems) and using the completeness of the eigenfunctions, the 4-point quantities are defined as
\small\begin{align}
S(\boldsymbol{r}_1,\boldsymbol{r}_2,\boldsymbol{r}_3,\boldsymbol{r}_4)=\hspace{-0.15cm}\sum_{\tiny n_1n_2 n_3n_4\normalsize}& S_{n_1n_2n_3n_4}\times\nonumber\\ &\psi_{n_1}(\boldsymbol{r}_1)\psi_{n_2}^{*}(\boldsymbol{r}_2)\psi_{n_3}^{*}(\boldsymbol{r}_3)\psi_{n_4}(\boldsymbol{r}_4),
\label{Stoeig}\end{align}\normalsize
where $S_{n_1n_2n_3n_4}=\int{\rm d}(\boldsymbol{r}_1...) S(\boldsymbol{r}_1,\boldsymbol{r}_2,\boldsymbol{r}_3,\boldsymbol{r}_4)\times$ $\psi_{n_1}^{*}(\boldsymbol{r}_1)\psi_{n_2}(\boldsymbol{r}_2)\psi_{n_3}(\boldsymbol{r}_3)\psi_{n_4}^{*}(\boldsymbol{r}_4)$, and the particle-particle propagator: $L^{pp}_{n_1n_2n_3n_4}=\int{\rm d}(\boldsymbol{r}_1...) L_{pp}(\boldsymbol{r}_1,\boldsymbol{r}_2,\boldsymbol{r}_3,\boldsymbol{r}_4)\times$ $\psi_{n_1}^{*}(\boldsymbol{r}_1)\psi_{n_2}^{*}(\boldsymbol{r}_2)\psi_{n_3}(\boldsymbol{r}_3)\psi_{n_4}(\boldsymbol{r}_4)$.

In Eq.~\ref{eq:Sigma_4p} the summations are over all indices. The summation(or rather integration) over wavevectors, however, will be dictated by the conservation of momentum, and the band indices are determined by the type of interaction being described (e.g., $n_1$ and $n_2$ for $\Pi$ must be particle-hole or hole-particle). This will become clearer below. If we look at the usual $GW$ term (without the exchange) and take only the first term in the expansion of $\Pi$, the self-energy in the frequency representation is

\begin{align}
\Sigma_{mn}^{GW}(\omega)=&\sum_{n',n_2\in o,n_1\in u}\frac{V_{\scriptsize\begin{array}{l}n'n\\n_1n_2\end{array}}V_{\scriptsize\begin{array}{l}n_1n_2\\n'm\end{array}}}{\omega-\varepsilon_{n'}-\varepsilon_{n_2}+\varepsilon_{n_1}-i\eta}+\nonumber\\
&\sum_{n',n_2\in u,n_1\in o}\frac{V_{\scriptsize\begin{array}{l}n'n\\n_1n_2\end{array}}V_{\scriptsize\begin{array}{l}n_1n_2\\n'm\end{array}}}{\omega-\varepsilon_{n'}-\varepsilon_{n_2}+\varepsilon_{n_1}+i\eta}+... .\label{eq:Sig0}
\end{align}

This diagram (and the second-order exchange equivalent) will be present in all channels and care will need to be taken to prevent multiple inclusions of this term.

The BSE for the irreducible polarisation is discussed in length in Ref.~\cite{QSGWhat}. The kernel there is simply $-W$ and it is easy to see that the kernel becomes $\widetilde{\chi}$ when instead calculating $\Pi$. To summarise from that work, the following expression is for the 4-point polarisation $\Pi$
\begin{equation}
\Pi_{\scriptsize\begin{array}{l}n_1n_2\\n_3n_4\end{array}}(\omega)=(f_{n_4}-f_{n_3})\sum_{\lambda}\displaystyle{\frac{X_{n_1n_2,\lambda}X_{n_3n_4,\lambda}^{*}}{\omega-E_\lambda+i(f_{n_2}-f_{n_1})\eta}},\label{eq:Pi_Herm}
\end{equation}
where $f_n$ are the single particle occupations, $f=1(0)$ for an occupied(unoccupied) state, i.e., a hole(particle), and $X_\lambda$ and $E_\lambda$ are the eigenvectors and eigenvalues of the following 2-particle matrix\footnote{For simplicity, $H$ is assumed Hermitian in Eq.~\ref{eq:Pi_Herm}. The case when the Tamm-Dancoff approximation (see text) is chosen.}
\begin{equation}
H_{\scriptsize\begin{array}{l}n_1n_2\\n_3n4\end{array}}=(\varepsilon_{n_1}-\varepsilon_{n_2})\delta_{n_1n_3}\delta_{n_2n_4}+(f_{n_2}-f_{n_1})\chi_{\scriptsize\begin{array}{l}n_1n_2\\n_3n_4\end{array}}.\label{eq:ph_mat}
\end{equation}
These expressions are derived from the BSE for the irreducible polarisation $\Pi=P_0+P_0\chi\Pi$, with
\begin{equation}
P^0_{\scriptsize\begin{array}{l}n_1n_2\\n_3n_4\end{array}}(\omega)=\frac{f_{n_2}-f_{n_1}}{\omega-(\varepsilon_{n_1}-\varepsilon_{n_2})+i(f_{n_2}-f_{n_1})\eta}\delta_{n_1n_3}\delta_{n_2n_4}.
\end{equation}
Note that $H$ has a $2\times 2$ block structure $\left(\begin{array}{cc}A&B\\-B^{*}&-A^{*}\end{array}\right)$. Ignoring the off-diagonal blocks suppresses coupling between negative and positive energy transitions and is refered to as the Tamm-Dancoff approximation.\cite{TD_myrta} It is clear that the energy eigenvalues of $H$ will come in positive and negative energy pairs.\footnote{The two terms in the expression for the $GW$ self energy are such that one will contain the positive energy eigenvalues and the other will contain the negative. It should also be noted that the sign of the eigenvalues depends on how $H$ is defined in terms of which block comes first, i.e., hole-particle or particle-hole.}
By a similar approach, the particle-particle(hole-hole) propagator is
\begin{equation}
L_{\scriptsize\begin{array}{l}n_1n_2\\n_3n4\end{array}}(\omega)=\mp\sum_\alpha\frac{Y_{n_1n_2,\alpha}Y_{n_3n_4,\alpha}^{*}}{\omega-\Omega_\alpha\pm i\eta},
\end{equation}
where the upper sign is for particles and the lower for holes, and $Y$ and $\Omega$ are the eigenvectors and eigenvalues of the following 2 particle matrix
\begin{equation}
H_{\scriptsize\begin{array}{l}n_1n_2\\n_3n4\end{array}}^{pp/hh}=(\varepsilon_{n_1}+\varepsilon_{n_2})\delta{n_1n_3}\delta_{n_2n_4}\pm W_{\scriptsize\begin{array}{l}n_2n_4\\n_3n_1\end{array}}.
\label{eq:pp_mat}
\end{equation}
Finally, the equation for the self-energy is
\begin{widetext}
\begin{align}
\Sigma_{mn}(\omega)=-\sum_{n'\in o}V_{\scriptsize\begin{array}{l}n'n\\n'm\end{array}}&+\sum_{n_1n_2}\left(\sum_{\scriptsize\begin{array}{c}n'n_4\in u\\n_3\in o\end{array}}\widetilde{\chi}_{\scriptsize\begin{array}{l}n'n\\n_1n_2\end{array}}\displaystyle{\frac{X_{n_1n_2,\lambda}X_{n_3n_4,\lambda}^{*}}{\omega-\varepsilon_{n'}+E_\lambda+i\eta}}V_{\scriptsize\begin{array}{l}n_3n_4\\n'm\end{array}}+\sum_{\scriptsize\begin{array}{c}n'n_4\in o\\n_3\in u\end{array}}\widetilde{\chi}_{\scriptsize\begin{array}{l}n'n\\n_1n_2\end{array}}\displaystyle{\frac{X_{n_1n_2,\lambda}X_{n_3n_4,\lambda}^{*}}{\omega-\varepsilon_{n'}+E_\lambda-i\eta}}V_{\scriptsize\begin{array}{l}n_3n_4\\n'm\end{array}}\right)\nonumber\\
&+\sum_{n_1n_2}\left(\sum_{\scriptsize\begin{array}{c}n_3n_4\in u\\n'\in o\end{array}}\chi_{\scriptsize\begin{array}{l}n_3n\\n'n_4\end{array}}^{pp}\displaystyle{\frac{Y_{n_1n_2,\alpha}Y_{n_3n_4,\alpha}^{*}}{\omega+\varepsilon_{n'}+\Omega_\alpha+i\eta}}V_{\scriptsize\begin{array}{l}n'n_2\\n_1m\end{array}}+\sum_{\scriptsize\begin{array}{c}n_3n_4\in o\\n'\in u\end{array}}\chi_{\scriptsize\begin{array}{l}n_3n\\n'n_4\end{array}}^{pp}\displaystyle{\frac{Y_{n_1n_2,\alpha}Y_{n_3n_4,\alpha}^{*}}{\omega+\varepsilon_{n'}-\Omega_\alpha-i\eta}}V_{\scriptsize\begin{array}{l}n'n_2\\n_1m\end{array}}\right)\nonumber\\
&+\sum_{n_1n_2}\left(\sum_{\scriptsize\begin{array}{c}n'n_4\in u\\n_3\in o\end{array}}\widetilde{\chi}_{\scriptsize\begin{array}{l}n_2n\\n_1n'\end{array}}^{ph}\displaystyle{\frac{X_{n_1n_2,\lambda}X_{n_3n_4,\lambda}^{*}}{\omega-\varepsilon_{n'}+E_\lambda+i\eta}}V_{\scriptsize\begin{array}{l}n_3n'\\n_4m\end{array}}+\sum_{\scriptsize\begin{array}{c}n'n_4\in o\\n_3\in u\end{array}}\widetilde{\chi}_{\scriptsize\begin{array}{l}n_2n\\n_1n'\end{array}}^{ph}\displaystyle{\frac{X_{n_1n_2,\lambda}X_{n_3n_4,\lambda}^{*}}{\omega-\varepsilon_{n'}+E_\lambda-i\eta}}V_{\scriptsize\begin{array}{l}n_3n'\\n_4m\end{array}}\right)\nonumber\\
&-2\Sigma_{mn}^0(\omega),\label{eq:finalEq}
\end{align}
\end{widetext}
where $\Sigma^0$ prevents the multiple inclusion of the zeroth order term in each channel and will be similar to that presented in Eq.~\ref{eq:Sig0}. To summarise, $\widetilde{\chi}=V-W$, $\widetilde{\chi}^{ph}=W-V$ and $\chi^{pp}=W-W$, where the first is a direct type of interaction and the second is exchange-like. Note that the $V$ in the pp and ph channel terms can be replaced with a $W$, since it was assumed $\Pi=-iGG$ (supressing RPA and vertex effects in the polarisation) for these terms and $v\Pi$ could have been replaced with $WP$ and the assumption $P=-iGG$ (supressing only vertex effects) could have been made instead.

The exchange term in Eq.~\ref{eq:finalEq} is exact and the next two terms with only the $V$ part from $\widetilde{\chi}$ produces the standard $GW$ terms, with the infinite series of interacting bubbles (RPA and vertex effects) in the polarisation (as in the QS$G\hat{W}$ method developed by the author in Ref.~\cite{QSGWhat}), but with the kernel in the BSE being $V-W$ instead of just $-W$, since the reducible polarisation is calcaulted instead of the irreducible one (avoiding the inversion of the dielectric matrix and the numerical frequency convolution). Including the $-W$ in $\widetilde{\chi}$ in Eq.~\ref{eq:finalEq} introduces vertex effects in the self-energy at the same level as that in the polarisation and leads to second order screened exchange, that further reduces the self-polarisation and introduces density matrix fluctuations\cite{dm_fluc}. We expect these to shift relatively flat bands in strongly correlated systems such as CuCl.\cite{QSGWhat,kutepov_cucl} Next we have the screened $T$-matrix diagrams that will include correlations between the added particles(holes) and the excited electrons(holes). The author investigated the $T$-matrix diagrams for a positron in isolated systems,\cite{nature_positron} where the correlated positron-electron diagrams (describing virtual positronium formation, whereby an electron tunnels and temporarily binds with the positron) were crucial for capturing certain physical processes such as positron binding. It is hoped that this work will instigate the inclusion of these beyond-$GW$ diagrams in electronic structure (and even antimatter) software packages.
\par
\section{Conclusions}
Analytic expressions for the many body self energy with a series of 2 body correlations is derived. This results in the usual exchange and $GW$ diagrams that can include RPA and excitonic effects. Vertex effects were also included ab-initio in the expression for the self energy $\Sigma=iGW\Gamma$, by explicitly calculating $\delta\Sigma/\delta G$. Terms that describe 3 mutually interacting bodies (2 particles[holes] and 1 hole[particle]) were neglected and the result was an inifite ladder series of interactions between the added particle(removed hole) and the excited electrons and holes; and second order exchange effects arose in all three channels. Second order vertex effects were also included to describe the infinte series of combinations between the added particle(removed hole) and the hole(excited-electron), analogous to the usual electron-hole RPA. The result is a series of equations that require the diagonalisation of two matrices: the usual particle-hole 2-particle matrix and a particle-particle and hole-hole matrix with similar structure. 

The calculation of the reducible (instead of the irreducible) polarisation means that the expensive numerical frequency integration methods - widely used to calculate the self-energy from the irreducible polarisation (see e.g., Ref.~\cite{QSGW_paper}) - are no longer required and numerical errors are removed. The RPA contribution to the self-energy is now calculated along with the BSE contribution at no real extra cost. However, the number of states/transitions that are considered explicitly in these matrices will be smaller than the number of states that can be included in the RPA (since only the 2-point polarisation is present in the RPA). As in Ref.~\cite{QSGWhat} transitions not included in the matrices can still be included at the level of the RPA. In that work it is discussed how the RPA contribution from a sufficiently large number of states is calclulated (since $W^{\rm RPA}$ is calculated and then used in the BSE) and then the contribution from the smaller subset of states (those considered in the BSE) is subtracted before being reincluded with excitonic effects present. The same approach can be taken with this method. The method can be iterated in a self-consistent framework, such as the QS$GW$ scheme,\cite{QSGW_paper,QSGWhat} which is a one-shot $GW$ method, but with the starting point (usually DFT) dependence removed. QS$GW$ has the benefit that the poles of $G$ and $G_0$ coincide and using $G_0$ (possibly with the inclusion of $Z$-factors) may be appropriate. Many implementations of QS$GW$ and its extensions\cite{QSGW_paper,QSGWhat,PhysRevB.103.L161104} rely on the Ward identity to some extent, whereby - for the long range component of the vertex - $G\Gamma\approx G_0$. The method presented includes $\Gamma$ and so one would have to investigate the validity of the substitution $G\rightarrow G_0$. Note that either $Z$-factors or the full $G$ (as used in figure~26 of Ref.~\cite{QSGWhat}) can be used in the expression for the self-energy and/or - as discsussed in the text - the equation for the vertex results in a series of terms that include the vertex and we can truncate these series by relying on the Ward identity here. 

Self-interaction (or rather self-polarisation/correlation) effects will be present in the method. Strongly correlated systems with many electrons occupying localised $d$- and $f$-states are becoming more widely studied and so it is important that we treat self-polarisation effects adequatly in these systems. The exchange will remove the self interaction in $H_0$, however, self-correlation errors will arise due to the added particles(holes) creating and interacting with excited-electrons/holes occupying the same quasiparticle level/band. Along with partial cancellation due to the direct and screened exchange terms in the $\chi$, and analysing Eq.~\ref{eq:finalEq}, it is clear that steps can be taken to remove a lot of the self correlation effects present. As suggested in Ref.~\cite{QSGWhat} with some modest extensions - such as those described in this work along with e.g., spin-fluctuations\cite{Stepanov19} (where $T$-matrix channels\cite{Friedrich21} and the particle-particle correlation function that governs superconductivity\cite{Acharyalafeas} are calculated) and electron-phonon\cite{savio24} interactions - that may be added hierarchically, a broadly applicable, high-fidelity ab initio approach to solving one- and two-particle properties of the many-body problem is within reach and this work should pave the way towards solving this problem. The extension to three-body correlated propagation\cite{stef,PhysRevLett.131.216401} is one obvious direction for extension of the method.
\begin{acknowledgments}
The author would like to thank Dermot Green (QUB) for his supervision during this work and all those involved in the {\it CCP flagship project: Quasiparticle Self-Consistent $GW$ for Next-Generation Electronic Structure}, especially Mark van Schilfgaarde (NREL, Golden, Co.) for his help and grateful for support from the Engineering and Physical Sciences Research Council, under grant EP/M011631/1. The author was funded by the European Research Council grant 804383 ‘ANTI-ATOM’.\end{acknowledgments}
\bibliography{references}
\end{document}